Gunnar Jacobsen*, Tianhua Xu, Sergei Popov, Sergey Sergeyev and Yimo Zhang

# Phase Noise Influence in Long-range Coherent Optical OFDM Systems with Delay Detection, IFFT Multiplexing and FFT Demodulation

**Abstract:** We present a study of the influence of dispersion induced phase noise for CO-OFDM systems using FFT multiplexing/IFFT demultiplexing techniques (software based). The software based system provides a method for a rigorous evaluation of the phase noise variance caused by Common Phase Error (CPE) and Inter-Carrier Interference (ICI) including – for the first time to our knowledge – in explicit form the effect of equalization enhanced phase noise (EEPN). This, in turns, leads to an analytic BER specification. Numerical results focus on a CO-OFDM system with 10–25 GS/s QPSK channel modulation. A worst case constellation configuration is identified for the phase noise influence and the resulting BER is compared to the BER of a conventional single channel QPSK system with the same capacity as the CO-OFDM implementation. Results are evaluated as a function of transmission distance. For both types of systems, the phase noise variance increases significantly with increasing transmission distance. For a total capacity of 400 (1000) Gbit/s, the transmission distance to have the BER < $10^{-2}$ for the worst case CO-OFDM design is less than 800 and 460 km, respectively, whereas for a single channel QPSK system it is less than 1400 and 560 km.

**PACS® (2010).** 42.25.Kb, 42.79.Sz

*Corresponding author: Gunnar Jacobsen: Acreo AB, Electrum 236, SE-16440, Kista, Sweden, E-mail: gunnar.jacobsen@acreo.se
**Tianhua Xu:** Acreo AB, Electrum 236, SE-16440, Kista, Sweden; Royal Institute of Technology, Stockholm, SE-16440, Sweden; Tianjin University, Tianjin 300072, P.R. China
**Sergei Popov:** Royal Institute of Technology, Stockholm, SE-16440, Sweden
**Sergey Sergeyev:** Aston University, Birmingham, B4 7ET, UK
**Yimo Zhang:** Tianjin University, Tianjin 300072, P.R. China

# 1 Introduction

High capacity coherent optical transmission research today has focus on achieving capacities in excess of 100 Gbit/s for transmission distances of 1000 km or more [1]. An essential part of the optical system design is the use of Discrete Signal Processing (DSP) techniques in both transmitter and receiver to eliminate costly hardware for dispersion compensation, polarization tracking and control, clock extraction, carrier phase extraction etc. such that there is no need to e.g. use optical injection locking which is difficult to implement in practical systems [2].

In the core part of the network, emphasis has been on long-range (high sensitivity) systems where coherent (homodyne) implementations of n-level Phase-Shift-Keying (nPSK) and Quadrature Amplitude Modulation (nQAM). Coherent 4PSK (QPSK) systems with in-phase and quadrature modulation and using polarization multiplexing have proven superior performance up to total bit rates in the order of 100 Gbit/s [1]. Due to practical limitations in the performance of digital-to-analogue (DA) and analogue-to-digital (AD) electronics which currently operates at maximum 56 Gbaud [3] higher baud-rates are difficult to achieve for QPSK systems. Higher system capacities can be obtained using higher constellations but at the expense of increased influence of additive optical noise and laser phase noise. Another alternative is multiplexing several QPSK modulated channels using sub-carrier multiplexing (SCM – see e.g. [4]) techniques or OFDM techniques. OFDM MUX/DEMUX techniques are seen as more spectrally efficient than SCM techniques. Only OFDM implementations will be further considered in this paper.

Optical coherent systems can be seen as a complementary technology to modern systems in the radio (mobile) domain. It is important to understand the differences in these implementations and these are mainly that the optical systems operate at significantly higher transmission speeds than their radio counterparts and that they use signal sources (transmitter and local oscillator lasers) which are significantly less coherent than radio sources. For nPSK and nQAM systems, DSP technology in the optical domain is entirely focused on high speed implementation of simple functions, such as AD/DA conversion. The use of high constellation transmission schemes is a way of lowering the DSP speed relative to the total capacity. Using OFDM as MUX/DEMUX technology is an alternative approach of very efficient lowering the DSP speed (per channel) and still maintaining 100 Gbit/s (or more) total system throughput. Coherent detection is considered for longer distance high capacity OFDM implementations [5]. The relatively low channel baud-rate leads to an influence of phase noise which can be more severe than for single channel systems [6] with low constellations where the system capacity per Hz can be traded against phase noise sensitivity [7, 8, 9].

Using nPSK or nQAM systems with DSP based dispersion compensation leads to strong influence of laser phase noise which is further enhanced by equalization enhanced phase noise (EEPN) originating from the local oscillator laser [10, 11]. OFDM systems with low per-channel capacity may use wrapping of the signal in the time domain (cyclic prefix) to account for dispersion effects in this way eliminating the need for DSP based compensation. However, it has to be noted that cyclic prefix only can be used to correct the intra OFDM-channel dispersion. Interchannel dispersion is insignificant for low channel Baudrates but for higher rates the interchannel dispersion requires DSP-type correction and EEPN will result from this even for OFDM systems [12, 13, 14]. Using an RF pilot carrier which is adjacent to or part of the OFDM channel grid is an effective method of eliminating the phase noise effect [15, 16, 18], but it has to be noted that the dispersion influenced delay of OFDM channels will make the elimination incomplete [18]. This leads to a transmission length dependent (dispersion enhanced) phase noise effect [18]. It is worth to mention that for nPSK and nQAM implementations the RF pilot carrier may eliminate the phase noise entirely. However, the EEPN cannot be eliminated [10, 17, 19]. For long distance OFDM implementations the RF pilot tone is not feasible. Then a system implementation with higher baud-rate per OFDM channel and delay-demodulation for each channel – as considered in this paper – is more practical in order to lower the phase noise influence.

System simulations (transmission experiments implemented in a software environment) have proven to be efficient design tools for nPSK/nQAM systems using partly university developed system models [17] and partly commercial simulation tools [20]. Such simulations for e.g. the bit-error-rate (BER) are possible because practical system implementations are now based on soft-decision forward-error-correction (FEC) where a "raw" BER (without FEC) of the order of $10^{-2}$ is sufficient [21]. For OFDM with tens or hundreds of signal channels, it is obvious that direct simulation of the OFDM system BER with independent simulation data (PRBS sequences) for each signal channel is a formidable task which is difficult for realization even for modern computers. Thus, it is of special interest for OFDM system models to develop insight based upon rigorous analytical models for important system parts.

## 2 System modeling and theory

Here we display layouts for CO-OFDM systems using IFFT MUX and FFT DEMUX in a software based system implementation (Fig. 1).

### 2.1 CO-OFDM system IFFT MUX and FFT DEMUX and detection

In the following, we will present the derivation for CO-OFDM systems employing IFFT MUX and FFT DEMUX and detection. During a symbol period $T$ the complex envelope (constellation position) of one of the $N$ transmitted OFDM signal (defined as shown in Fig. 1) is $a_k$ ($k = 0, 1, \ldots, N-1$). Symbol of number $k$ is moved to the electrical carrier frequency $f_k = k/T$. The $N$ symbols are multiplexed (added) using IFFT, and the multiplexed signal is denoted $A(t) \cdot \exp(j(\varphi(t)))$. The multiplexed signal is put onto the optical carrier wave and the resulting signal in the optical domain is:

$$s(t) \equiv A(t) \cdot \exp\left(j\left(2\pi f_o t + \psi_{Tx}(t) + \varphi(t)\right)\right)$$
$$= e^{j(2\pi f_o t + \psi_{Tx}(t))} \sum_{k=0}^{N-1} a_k e^{j2\pi(k/N)m} \quad (1)$$

where the sampled time is defined (modulo $T$) in the interval $mN < t/T < (m+1)N$ with $0 < m < N-1$, $\psi_{Tx}(t)$ denotes the $Tx$ laser phase noise and $f_o$ the optical carrier frequency. The electrically multiplexed signal is the analogue output after digital Inverse Fast Fourier Transform (IFFT) of the digitized input sampled with $N$ bins separated by $T/N$, and each sample specifying one OFDM channel constellation $a_k$. After coherent detection with a local oscillator (LO) laser with the same carrier frequency as the $Tx$ laser, the output of the receiver, including FFT demodulation, chromatic dispersion compensation, cyclic

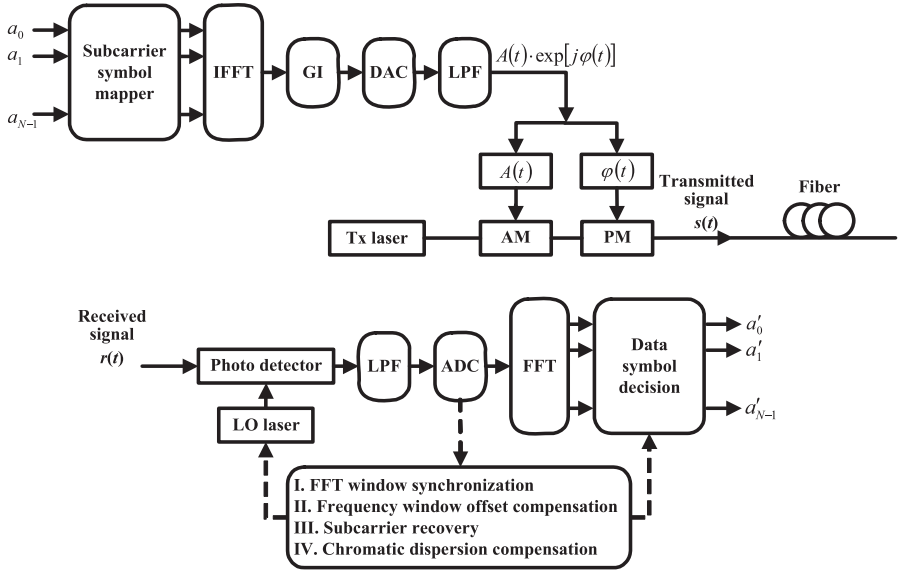

**Fig. 1:** OFDM system including IFFT MUX and FFT with an RF pilot tone for phase noise mitigation. The mathematics for the MUX and DEMUX is schematically indicated and discussed in detail in the text. Figure abbreviations: $a_0$–$a_{N-1}$ – constellation of $N$ transmitted OFDM symbols; $a'_0$–$a'_{N-1}$ – constellation of $N$ received OFDM symbols; IFFT – Inverse Fast Fourier Transform; GI – guard time insertion; DAC – discrete to analogue conversion; LPF – low pass filter; AM – amplitude modulator; PM – phase modulator, Tx – transmitter, LO – local oscillator; RF – radio frequency; ADC – analogue to discrete conversion; FFT – Fast Fourier Transform.

prefix correction and correlation detection the result is for symbol $k$ ($0 \leq k \leq N-1$) [1]:

$$a'_k = \frac{e^{-j(2\pi f_o t + \psi_{LO}(t))}}{N} \sum_{m=0}^{N-1} s(t) e^{-j2\pi(k/N)m} \quad (2)$$

where $\psi_{LO}(t)$ denotes the LO laser phase noise (including the equalization enhanced phase noise to be discussed in detail later [10, 17, 18, 19]). In the case of no phase noise influence, the orthogonality between the channels means that $a'_k = a_k$ and the symbol detection is perfect. Taylor expansion is now employed to identify the leading order phase noise influence in (2). The resulting Common Phase Error (CPE) for channel $k$ is:

$$\frac{j}{N} \sum_{m=0}^{N-1} \psi\left(\frac{mT}{N}\right) \quad (3)$$

The Inter-Carrier Interference (ICI) is:

$$\frac{j}{N} \sum_{\substack{r=0 \\ r \neq k}}^{N-1} a_r \sum_{m=0}^{N-1} \psi\left(\frac{mT}{N}\right) \cdot \exp\left(\frac{j2\pi(r-k)m}{N}\right) \quad (4)$$

It is possible to derive the phase noise variance in exact form accounting in detail for the partial phase noise correlation between different channel locations in the OFDM frame. This can be done by introducing the correlation coefficient between two time-overlaping Wiener processes specified eg. By $m = s$ and $m = r$. They have the correlation coefficient $\rho_{p,q} = (1 - |p - q|/N)^{1/2}$ with $\rho_{p,p} = \rho_{q,q} = 1$. Each phase noise sample $\psi(mT/N)$ is sampled once per symbol time $T$ (i.e. specifies the phase noise evolution over $T$) and is therefore given by a Wiener process with zero-mean Gaussian probability density function (pdf) with variance $\sigma^2$ ($\sigma^2$ will be specified later in this section). Then the variance of (3) and (4) (which needs to be considered together due to the effects of the partial correlation) is given by

$$\sigma^2_{k,CPE+ICI} = \frac{\sigma^2}{N^2} \sum_{q=0}^{N-1} \left( \sum_{p=0}^{N-1} \left[ \sum_{r=0}^{N-1} \text{Re}\left( \frac{a_r}{a_k} \exp\left( j\frac{2\pi(r-k)q}{N} \right) \right) \right] \right.$$
$$\left. \cdot \rho_{p,q} \cdot \left[ \sum_{s=0}^{N-1} \text{Re}\left( \frac{a_s}{a_k} \exp\left( j\frac{2\pi(s-k)p}{N} \right) \right) \right] \right) \quad (5)$$

We note that the time correlation between contributions from neighboring channels is strong ($\rho_{p,q} \approx 1$ in this case).

For the final demodulation of one OFDM channel operating as a 10–25 GS/s QPSK system, we have to consider electronic CD compensation (correcting the inter-channel dispersion). In this case, the phase noise variance $\sigma^2$ is influenced by EEPN, and it is given as [17]:

$$\sigma^2 = 2\pi(\Delta\nu_{Tx} + \Delta\nu_{LO}) \cdot T + \frac{\pi\lambda^2}{2c} \cdot \frac{D \cdot L \cdot \Delta\nu_{LO}}{T}$$
$$\equiv 2\pi(\Delta\nu_{Tx} + \Delta\nu_{LO} + \Delta\nu_{EEPN})T \quad (6)$$

where $T$ is the symbol time, $D$ is the fiber dispersion coefficient, $c$ is the free space velocity of light, $\lambda$ is the wavelength, $L$ is the fiber length and transmitter and local Oscillator linewidths are denoted $\Delta v_{Tx}$ and $\Delta v_{LO}$. In (5)–(6) it is observed that the intra-channel dispersion is corrected by using cyclic prefix whereas the inter-channel dispersion needs electronic dispersion compensation and thus is subject to EEPN. For a single-channel QPSK system with the same capacity as an $N$-channel OFDM system (6) describes the phase noise influence provided that the symbol time is adjusted to the QPSK bit-rate, i.e. the resulting symbol time is $T/N$. This indicates that the resulting EEPN effect is significantly more pronounced for the single channel high capacity QPSK system.

We will investigate the resulting phase noise variance in more detail in the numerical examples of the next section.

When considering the amplitude of the phase noise contribution (for OFDM systems) which influences detection of the length (magnitude) of $a_k$, there is no contribution from the CPE part of the phase noise as can be seen from (4). The ICI part will give a contribution (from the real part of (5)) which can be specified in similar forms as (6).

We note that practical nPSK, as well as nQAM, systems can be designed by choosing constellation configurations such that the phase noise influence on the detected phase is the dominating phase noise contribution. In the following, we will not consider the magnitude part of the phase noise influence.

## 3 Simulation results and discussion

It is of interest to compare the normalized (dividing by the intrinsic phase noise variance $\sigma^2$) CPE + ICI phase noise influence in (5). With this normalization we will observe the phase noise influence relative to that of a single channel QPSK system with bit-rate $1/T$. We consider an OFDM system implementation with 4PSK (QPSK) channel modulation.

It is appropriate to evaluate (5) for all combinations of constellations between the OFDM channels (considering for QPSK channel modulation 4 different constellations per channel) and for all demodulated channels (for all $k$-values). We note that for $N$ OFDM channels this leads to an evaluation of $N \cdot 4^N$ cases for a full investigation and this quickly renders the practical evaluation impossible for increasing $N$.

Fig. 2 shows the results as a function of the number of OFDM channels, $N$, for a received OFDM frame where all symbols $a_r$ ($r = 0, 1, \ldots, N – 1$) are the same, and results are shown for the received channel number 0 ($k = 0$). Results are evaluated using full time correlation between phase noise samples (using (5) and defining the correlation coefficient $\rho_{s,m} = 1$ for all $s$ and $m$ values) and for partial correlation using (5). For an $N$-channel CO-OFDM system it is of interest to note that the normalized worst case influence (on the variance) is $N$ both in the case of full and partial correlation.

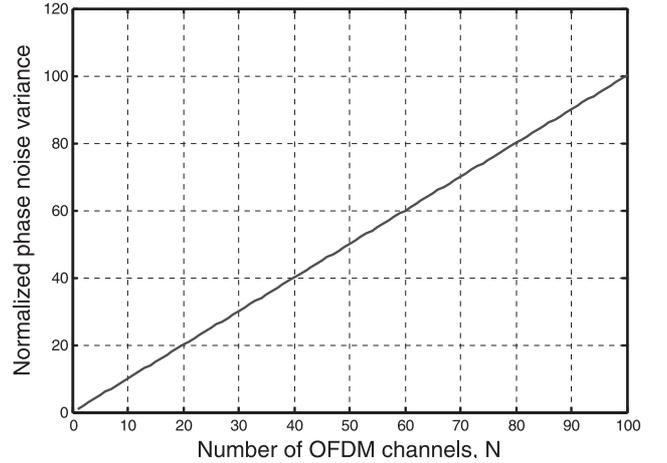

**Fig. 2:** Normalized phase noise variance $\sigma^2_{k,CPE+ICI} / \sigma^2$ as a function of the number of OFDM channels $N$ for received channel $k = 0, N – 1$. Dashed, full and dotted curves shows results in the cases of full, partial and no time-correlation between phase noise using (6).

We will investigate the validity of the results in Fig. 2 in some detail. We evaluate the normalized phase noise variance for all constellation configurations and all received channel positions in the OFDM grid for the most important practical design case – the partly fully correlated case considered in (5). We do that for $N = 2, 3, \ldots, 9$ and display representative results for $N = 5, 9$ in Fig. 3 in bar diagram format. From Fig. 3 it is clearly observed that system design based on a normalized phase noise variance of $N$ (as used in Fig. 2) represents a sensible worst case for the selected $N$-values. We tentatively extract this observation to cover all larger $N$-values as well (where the results of Fig. 3 cannot be generated due to the huge amount of $N \cdot 4^N$ required evaluation cases) and also assume – in accordance with the results of Fig. 3 – that normalized phase noise variance of $N$ is reasonable as a worst case system design scenario. In Fig. 3 it is obvious that the normalized phase noise is mostly close to zero and the nonzero part where it is larger than 0.1 – say – corresponds to less than a fraction of $10^{-2}$ of the total number of possible constellation combinations. This means that the phase noise influence is largely eliminated in the OFDM receiver and it might lead to very optimistic phase noise design criteria for the CO-OFDM system at hand. However, it must be remembered that this conclusion is based upon a leading order Taylor expansion of the phase noise influence (see (2)–(3)). This leading order approximation is increasingly inaccurate for larger phase noise values and in the following we will base our results on the worst case assumption that the normalized phase noise variance takes a value in the order of $N$. Here the leading order Taylor expansion is expected to be reasonable, but it should be noted that detailed verification of when this is the case is an important subject for future CO-OFDM system research.

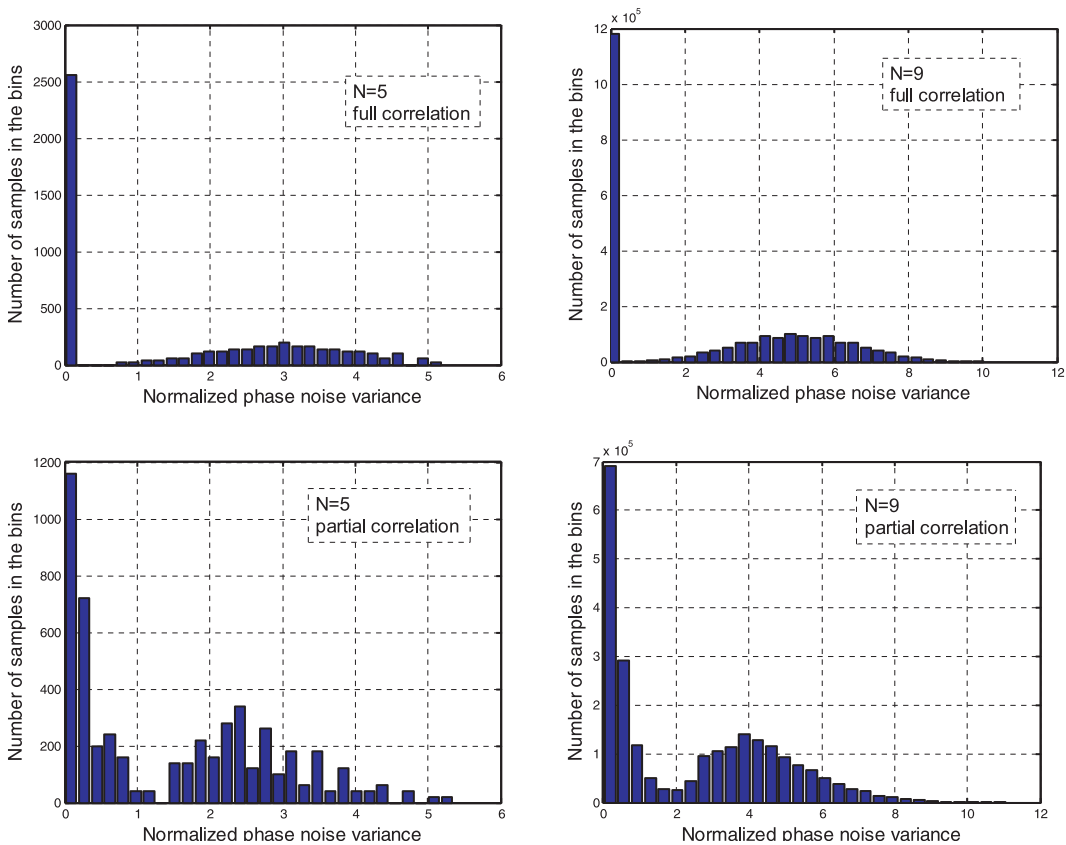

**Fig. 3:** Number of samples in a bin representation versus normalized phase noise variance $\sigma^2_{k,CPE+ICI}/\sigma^2$ using (5) in the case of partial and full time-correlation between phase noise samples from neighboring channels (as indicated). Number of OFDM channels considered are $N = 5$ and 9 (as indicated) and all constellation configurations and all received channels are considered.

We will now move to more detailed practical CO-OFDM system examples. We consider a normal transmission fiber ($D = 16$ ps/nm/km) for the distances up to around 2000 km, a transmission wavelength of $\lambda = 1.55$ μm, $c = 3 \cdot 10^8$ m/s, an OFDM channel separation of $\Delta f = 10$ GHz, i.e. baud rate 10 (25) GS/s (symbol time $T = 0.1$ (0.04) ns), channel modulation as QPSK, and the number of channels $N$ of 10. This gives a total OFDM system capacity in a dual polarization implementation of 400 (1000) Gbit/s. We also consider a single channel QPSK system with baud rates of 100 GS/s (representing the upper limit for current research implementations) and 250 GS/s (representing a possible future advanced QPSK system). We note that a system capacity of >250 GS/s is in principle feasible in the CO-OFDM configuration with today's technology by adding more 25 GS/s OFDM channels.

The phase noise parameter of interest is specified by (5) and (6), and may, in general form, be denoted $\sigma$. The BER floor for the two system implementations is given as [6]:

$$BER_{floor} \approx \frac{1}{2} erfc\left(\frac{\pi}{4\sqrt{2}\sigma}\right) \quad (11)$$

In Figure 4, we display the $BER_{floor}$ versus transmission distance. A reasonable practical system design constraint is that the BER floor should be below $10^{-2}$ in order for soft Forward Error Correction (FEC) techniqes to operate well [20]. It can be seen that the OFDM systems with capacities of 400, and 1000 Gbit/s fulfill this requirement for $L < 800$ and 460 km. The distance for single channel QPSK systems with 400 and 1000 Gbit/s capacity is 1400 and 560 km. Thus the worst case OFDM performance is slightly poorer than the performance of a single cahnnel QPSK system with the same capacity.

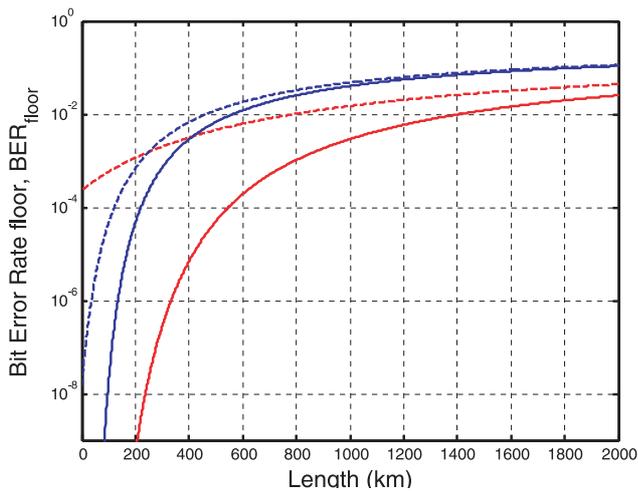

**Fig. 4:** Bit-error-rate floor as a function of transmission length for QPSK and 10 channel OFDM systems with total capacities of 100 GS/s (red curves) and 250 GS/s (blue curves). OFDM (QPSK) system performance is shown by dashed (solid) curves. Dashed curves are for 250 GS/s QPSK and 10 channel (25 GS/s per channel) OFDM systems. Solid curves are for 100 GS/s QPSK and 10 channel (10 GS/s per channel) OFDM systems.

## 4 Conclusions

We present a comparative study of the influence of dispersion for CO-OFDM systems influenced by equalization enhanced phase noise using software based FFT multiplexing and IFFT demultiplexing techniques. This is, to our knowledge, the first detailed and rigorous study of this OFDM system configuration. From the analysis it appears that the phase noise influence for the two OFDM implementations is similar. It can be also seen that the theoretical formulation for the software based system provides a method for a rigorous evaluation of the phase noise variance caused by Common Phase Error (CPE) and Inter-Carrier Interference (ICI), and this, in turns, leads to a BER specification.

A major novel theoretical result specifies in exact form the resulting phase noise variance accounting for the combined CPE and ICI influence including the partial correlation between ICI phase noise samples of different OFDM channels. From a statistical analysis we have used the formulation to identify the worst case phase noise influence in the OFDM system with QPSK channel modulation. The worst case value has been used in the system design and it has been found that the OFDM worst case implementation performs slightly worse than a single channel QPSK system with the same capacity.

The numerical results of the current study focus on a worst case specification for a CO-OFDM system with 10 and 25 GS/s QPSK channel modulation and 100 and 250 GS/s total system capacity. BER results are evaluated and compared to the BER of a single channel QPSK system of the same capacity as the OFDM implementation. A system capacity of 250 GS/s cannot be realized with current digital to analogue or analogue to digital (DA/AD) circuits whereas a system capacity of >250 GS/s is in principle feasible in the CO-OFDM configuration with today's technology by adding more 25 GS/s OFDM channels.

Results are evaluated as a function of transmission distance. The influence of equalization enhanced phase noise (EEPN) is included. For both type of systems, the phase noise variance increases very much with increasing the transmission distance and the two types of systems have closely the same BER as a function of transmission distance for the same capacity. For the 100 (250) GS/s the transmission distance to have the BER $< 10^{-2}$ is less than 800 and 460 km, respectively. The distance for single channel QPSK systems with 100 and 250 Gbit/s capacity is 1400 and 560 km.